\documentstyle[a4,12pt]{article}
\def\be{\begin{equation}}
\def\ee{\end{equation}}
\def\beq{\begin{eqnarray}}
\def\eeq{\end{eqnarray}}

\def\gsim{\:\raisebox{-0.5ex}{$\stackrel{\textstyle>}{\sim}$}\:}

\newcommand{\ie}{{\it ie}}

\newcommand{\etal}{{\it et al.}}
\newcommand{\gev}{{\rm GeV}}
\newcommand{\fm}{{\rm fm}}
\newcommand{\order}[1]{${\cal O}(#1)$}
\newcommand{\morder}[1]{{\cal O}(#1)}
\newcommand{\eq}[1]{Eq.\ (\ref{#1})}
\newcommand{\lqcd}{\Lambda_{QCD}}
\newcommand{\ave}[1]{\langle{#1}\rangle}
\newcommand{\egl}{\varepsilon_g}
\newcommand{\equ}{\varepsilon_q}
\newcommand{\mperpi}{m_{\perp i}}

\begin{document}
\begin{flushright}
TIFR/TH/97-15 \\
NORDITA--97/31 P\\
hep-ph/9705273\\
\today
\end{flushright}
\bigskip
\begin{center}
{\Large{\bf The Intrinsic Gluon Component of the Nucleon}} \\[3cm]
{\large Paul Hoyer$^1$ and D.P. Roy$^2$} \\[1cm]
$^1$ Nordita, Blegdamsvej 17, DK-2100 Copenhagen, Denmark \\ [2mm]
$^2$ T.I.F.R., Homi Bhabha Road, Mumbai 400 005, India.
\end{center}
\bigskip\bigskip
\begin{center}
\underbar{\bf Abstract}
\end{center}
\medskip
Using an intrinsic parton model
we estimate the rough shape and size of
the intrinsic gluon component of the nucleon, corresponding to an
energy scale $Q$ of the order $\lqcd$.
It is nearly as hard in shape
as the valence quark, while its size accounts for a quarter of the
nucleon momentum. Both are in qualitative agreement with the input
leading twist
gluon distribution assumed by Gl\"uck,
Reya and Vogt at this scale in
order to reproduce the observed distributions at $Q^2 \gsim 1$ GeV$^2$
via perturbative QCD evolution.

\newpage

While the $Q^2$ dependence of the parton distribution functions are
successfully explained
by perturbative QCD,
one cannot predict the shape and size
of these distribution functions at the starting
scale of $Q^2$. The standard practice is to use phenomenological
parametrisations of the quark and gluon distributions in the Bjorken
$x$ variable [1] at an input scale of
\be
Q^2_0 = 1 - 2\ \gev^2 . \label{highqodef}
\ee
Roughly speaking, in this $Q^2$ region the nucleon momentum seems to
be shared about equally between the three valence quarks and an
infinite number of soft gluons, with the gluon distribution function
\be
g(x) \propto x^{-1} \ {\rm at\ low} \ x  . \label{softgluedist}
\ee
Thus to a first approximation
\beq
\egl &\equiv& \int^1_0 x g (x) \ dx \simeq {1 \over 2} \ ,
\label{softegfract} \\ [3mm]
\bar x_g &\equiv& \frac{\egl}{\int^1_0 g (x) \, dx} = 0 ,
\label{softxgfract}
\eeq
where $\egl$ is the
total fraction of the nucleon momentum which is carried by gluons
and $\bar x_g$ is the average momentum
fraction carried per gluon. There is so far no theoretical
understanding of the size or
the shape of the input gluon
distribution represented by
Eqs. (\ref{softegfract}), (\ref{softxgfract})
even at a qualitative level.

A bold step along this direction was taken by Gl\"uck, Reya and Vogt
(GRV)
[2, 3, 4] by extending the perturbative QCD evolution of the
leading twist parton distribution functions down to
\be
Q^2_0 \simeq 0.2 \ \gev^2 \ ,  \label{lowqodef}
\ee
\ie, to a $Q_0$ of the same order as $\lqcd$. This corresponds to the
regime of long distance and time scales, where the input parton
distributions can be regarded as intrinsic to the nucleon [5].
An initial attempt to generate the canonical gluon distribution
represented by
(\ref{highqodef}) - (\ref{softxgfract})
by starting only with the three valence quarks at
the low
input scale of (\ref{lowqodef}) did not succeed [2].
However, GRV did reproduce the measured
gluon distribution by adding a valence-like
gluon component to the valence quarks at the
low input scale.
While the shape of the input gluon
distribution is roughly similar to that of the valence quarks,
the gluons carry a smaller momentum fraction
\be
\egl = 0.25 \ \ {\rm and} \ \ \equ = 0.60 \ \ {\rm at} \ \ Q^2_0 \simeq
0.2\ \gev^2 ,  \label{hardfract}
\ee
in the case of leading order QCD evolution [4]. The remaining 15\% of
the nucleon momentum is attributed to a
sea quark component whose
shape is also assumed to be valence like.

In this note we shall provide rough estimates of the shape and
size of the intrinsic gluon component, corresponding to the long time
scale of \eq{lowqodef}
above. For this purpose we shall use the model of
Brodsky \etal\ [5] for the long time scale structure
of the nucleon
wavefunction. As we shall see below, the predicted intrinsic gluon
component has a roughly similar shape and size as the above mentioned
inputs of GRV [2-4].

According to the model of [5], the partons in the nucleon Fock states
\be
|qqq\rangle \:, \ |qqqg\rangle \:, \ |qqq\bar q q\rangle \:, \cdots
\label{fockdecomp}
\ee
have similar
velocities in order to stay together over a long time scale. More
quantitatively, the
probability $P_n$ of an $n$-parton Fock state is given by
\be
P_n \propto (\Delta E)^{-2} \propto
\left(m_N^2-\sum^n_{i=1} \ {\mperpi^2 \over
x_i}\right)^{-2} \ .  \label{fockprob}
\ee
Here $\Delta E$ is the energy difference, in the infinite momentum frame,
between the nucleon and the Fock state, and
$\mperpi^2 = m^2_i + k^2_{\perp i}$
is the squared transverse mass of parton $i$.
The distribution (\ref{fockprob}),
motivated by old fashioned perturbation theory [5],
is relevant for the long time-scale
$(\propto 1/\Delta E)$ structure of the nucleon. The leading twist $Q^2$
evolution, on the other hand, reflects the increasing resolution of
short-lived `extrinsic' states created through single parton splitting. In
the spirit of GRV we thus propose using \eq{fockprob} to determine the input,
low $Q^2$ `valence' distribution to which leading twist evolution is
applied.

According to \eq{fockprob} the probability distribution of a given
Fock state is peaked at
\be
x_i = \frac{\mperpi}{\sum_i \ \mperpi} \ .  \label{peakprob}
\ee
In particular,
if there are heavy partons in a Fock state then they
will carry a
large fraction of the nucleon momentum. This led to the
suggestion of an intrinsic charm component $|qqq\bar c c\rangle$ of the
nucleon, where the charm quark pair carries
the bulk of the nucleon momentum [5]. The EMC data on muon induced dimuons
[6] seems to indicate the presence of such a hard intrinsic charm component
in the nucleon [7],
but there is no definitive experimental evidence
for it so far.  For fixed $\mperpi$,
the probability
distribution of \eq{fockprob} implies a
power-law fall-off
$(1-x)^n$ for the parton distributions, with $n=3$ and 4 for the
valence quark and gluon respectively [5, 8]. A similar model was
used in [9] to predict a hard fragmentation function for the charm
quark into a charmed hadron, which is in good agreement with
experimental data.

In the present case we are interested in the long time scale structure
of the nucleon wavefunction in terms of
the multi-gluon Fock states
\be
|qqq\rangle \:, \ |qqqg\rangle \:, \ |qqqgg\rangle \:, \cdots .
\ee
Consider the $n$-parton Fock state consisting of the 3 valence quarks
and ($n-3$) gluons.
We take the parton momenta to be
distributed according to
\eq{fockprob} with a common transverse mass
\be
\mperpi \simeq \langle k^2_\perp \rangle^{1\over 2} \simeq 0.3 - 0.4\ \gev ,
\label{kvalue}
\ee
\ie, with a
typical intrinsic momentum corresponding to a hadronic scale
of \order{1\ \fm}. This implies
an equipartition of the
nucleon momentum among the $n$ partons, \ie,
\be
\bar x_q = \bar x_g = 1/n \ , \ \ \equ = 3/n \ , \ \ \egl = (n-3)/n .
\label{fockfrac}
\ee
Thus if a single Fock state were dominant the shape of the gluon
distribution would be identical to that of
the valence quark. However, in general
we have to consider the contribution of all nucleon
Fock states
\be
|N\rangle = A_3 |qqq\rangle + A_4 |qqqg\rangle + A_5 |qqqgg\rangle + \cdots .
\label{focksum}
\ee
Here
\be
C_n = |A_n|^2 = \int\, P_n (x_1 , \ldots , x_n) \prod^n_{i=1} \ dx_i
\label{cndef}
\ee
represents the net probability for the $n$-parton Fock state, and
$\sum_{n=3} C_n = 1$. Thus
\beq
\equ &=& \sum_{n=3} {3 C_n \over n} = 3 \left\langle{1\over
n}\right\rangle = 3\bar x_q \nonumber \\ [3mm]
\egl &=& \sum_{n=3} {(n-3) C_n \over n} =
\left\langle{n-3\over n}\right\rangle ,  \nonumber \\
\bar x_g &=& \frac{\egl}{\ave{n_g}} =
\left\langle{n-3\over n}\right\rangle/\ave{n-3} ,
\label{fracexpr}
\eeq
\ie, in general $\bar x_g < \bar x_q$. As we shall see below, the
intrinsic gluon distribution can be soft $(\bar x_g \simeq 0)$ or
nearly as hard as the valence quark $(\bar x_g \sim \bar x_q)$
depending on the nature of the coefficients $C_n$.

In order to proceed further we need to know the $n$-dependence of the
probability factors $C_n$ of Eqs. (\ref{fockprob}) and (\ref{cndef}).
We shall assume that the $n$-dependence is mainly determined by
the energy denominators of \eq{fockprob}
evaluated at their most likely configuration (\ref{peakprob}), with
$\mperpi$ given by \eq{kvalue}. This gives (neglecting the nucleon mass term
in \eq{fockprob}),
\be
C_n \propto 1/n^4 .  \label{cn4}
\ee
Consequently,
\be
\bar x_q = \int^{\infty}_3 {dn\over n^5} \left/ \int^{\infty}_3 {dn
\over n^4} \right. = {1 \over 4} \ , \ \ \equ = {3 \over 4} ,
\label{qfrac4}
\ee
\be
\egl = {1 \over 4} \ , \ \ave{n_g} = {3\over 2} \ , \ \
\bar x_g = \frac{\egl}{\ave{n_g}} = {1\over 6} .
\label{gfrac4}
\ee
Thus the average momentum fractions of gluons and quarks are similar
$(\bar x_g \simeq 0.7 \bar x_q)$, while their total momentum
fractions are in the ratio $\egl \colon \equ = 1 \colon 3$. Both
features are in qualitative agreement with the input gluon
distribution of GRV at $Q^2_0 \simeq 0.2\ \gev^2$ [4], as discussed
above.

It is instructive to see how sensitive the results are to the assumed
$n$-dependence of the probability factors $C_n$.
Let us consider the three other cases
\be
C_n \propto 1/n \ , \ \ 1/n^2 \ , \ \ {\rm and} \ \ 1/n^3 \ .
\ee
The resulting average and total momentum fractions as well as $\ave{n_g}$
are shown
in Table I along with those of Eqs. (\ref{cn4}) - (\ref{gfrac4}).
We see that the shape and size
of the intrinsic gluon distribution depend sensitively on
the distribution of Fock states. While both the quark and the gluon
distributions are hard for $C_n \propto 1/n^4$, they are both soft for
$C_n \propto 1/n$. Simultaneously the
total momentum fraction
carried by the gluons $(\egl)$ increases from 1/4 to 1.

It may also be noticed from Table I that
$C_n \propto 1/n^2$ corresponds to hard quark and soft gluon
distributions, each carrying half the nucleon momentum fraction, as in
the case of the canonical parametrisation at higher $Q^2$ given by
Eqs. (\ref{highqodef}) - (\ref{softxgfract}).
However, in this perturbative regime
the virtual photon scatters from Fock states having a short life-time of
\order{1/Q}. Hence the dynamics is not determined by the Fock states of
lowest energy, as in our intrinsic model based on \eq{fockprob}.

Given that the parton distributions depend
sensitively on the Fock state probabilities, we find it
significant that the probability distribution (\ref{cn4}) of intrinsic states
gives roughly the right shape and size of the valence parton distributions,
as required for the GRV input at the appropriate scale of $Q_0^2 \simeq 0.2\
\gev^2$ [2-4]. It would be interesting to compare the Fock state
distribution of \eq{cn4} with that of solvable field theory models, such as
QCD$_{1+1}$ [10] and dimensionally reduced QCD [11].

So far we have neglected the (presumably small) intrinsic sea quark
component. Regardless of our specific model, there are at least
two reasons to expect that the sea
quark component should, like the gluon, have a hard distribution
at the low momentum scale of \eq{lowqodef},
as is indeed the case in the GRV input [4].
(i) The presence of soft sea quarks $(x_i \rightarrow 0)$ in any Fock
state will imply the corresponding $\Delta E \rightarrow \infty$,
making such states irrelevant for the long time scale structure.
(ii) The
perturbative evolution of sea quarks in the small $x$ region is driven
by the small $x$ behavior of gluons. Thus a soft sea component cannot
develop as long as the gluon component remains hard.

It should be emphasized that the hard quark and gluon distributions at the
low scale of \eq{lowqodef} represent leading twist parton distributions
which are not directly measurable in electron scattering. At such low values
of $Q^2$ the physical cross section is in fact dominated by higher twist
contributions -- the photon scatters coherently from several quarks. Being
of leading twist, the $Q^2$ evolution of the parton distributions are
known, however, and they can thus be compared with data at a higher scale,
such as that given by (\ref{highqodef}). This was, of course, how GRV
arrived at their parametrization of the parton distributions at low $Q^2$.

In summary, we have estimated the rough shape and size of the
intrinsic valence quark and gluon components
of the nucleon. At the low scale $Q^2 \simeq 0.2\ \gev^2$, \ie, for $Q=
\morder{\lqcd}$, we assume the nucleon Fock state probabilities to be
proportional to
$(\Delta E)^{-2}$, where $\Delta E$ is the excitation energy of the state.
In this approach the average momentum of an intrinsic gluon turns out to be
similar to that of a valence quark. The total momentum fractions carried
by gluons and quarks are in the ratio 1 : 3. Both features are in qualitative
agreement with the shape and size of the input parton distributions found
by GRV [2--4] at $Q^2 \simeq 0.2\ \gev^2$. When evolved to $Q^2 \gsim 1\
\gev^2$ these distributions reproduce the experimental data. Thus our model,
taken together with the GRV analysis, provides a theoretical basis for
understanding the shape and size of the observed gluon distribution at $Q^2
= 1-2\ \gev^2$.

It is a pleasure to thank Profs. S. Brodsky, R.M. Godbole, J. Kwiecinski, E.
Reya, R.G. Roberts and G.G. Ross for illuminating discussions.

\vskip 3em

\noindent\underbar{\bf REFERENCES}

\begin{enumerate}
\item[{1.}] See e.g. A.D. Martin, R.G. Roberts and W.J. Stirling, {\it
Phys. Lett.} {\bf B 387} (1996) 419, hep-ph/9606345.
\item[{2.}] M. Gl\"uck, E. Reya and A. Vogt, {\it Z. Phys.} {\bf C 48}
(1990) 471.
\item[{3.}] M. Gl\"uck, E. Reya and A. Vogt, {\it Z. Phys.} {\bf C 53}
(1992) 127; {\it Phys. Lett.} {\bf B 306} (1993) 391.
\item[{4.}] M. Gl\"uck, E. Reya and A. Vogt, {\it Z. Phys} {\bf C 67}
(1995) 433.
\item[{5.}] S.J. Brodsky, P. Hoyer, C. Peterson and N. Sakai, {\it
Phys. Lett.} {\bf B 93} (1980) 451; S.J. Brodsky, C. Peterson and N.
Sakai, {\it Phys. Rev.} {\bf D 23} (1981) 2745.
\item[{6.}] EMC Collaboration: J.J. Aubert et. al, {\it Nucl. Phys.}
{\bf B 213} (1983) 31.
\item[{7.}] D.P. Roy, On the Indication of Hard Charm in the Latest
EMC Dimuon Data, TIFR/TH/83-1 (1983); see also B. Harris, J. Smith and
R. Vogt, {\it Nucl. Phys.} {\bf B 461} (1996) 181, hep-ph/9508403.
\item[{8.}] S.J. Brodsky and I. Schmidt, {\it Phys. Lett.} {\bf B
234} (1990) 144; S.J. Brodsky, M. Burkhardt and I. Schmidt, {\it Nucl.
Phys.} {\bf B 441} (1995) 197, hep-ph/9401328.
\item[{9.}] M. Suzuki, {\it Phys. Lett.} {\bf B 71} (1977) 139; see
also J.D. Bjorken, {\it Phys. Rev.} {\bf D 17} (1978) 171.
\item[{10.}] K. Hornbostel, S. J. Brodsky and H. C.
Pauli, {\it Phys. Rev.} {\bf D 41} (1990) 3814.
\item[{11.}] F. Antonuccio and S. Dalley, {\it Phys. Lett.} {\bf B376} (1996)
154, hep-th/9512106.
\end{enumerate}

\newpage

\noindent{\bf Table I}: The average and total momentum fractions carried by
the valence quarks and the intrinsic gluons are shown along with the average
number of gluons for four different types of Fock state distributions.

\begin{center}
\begin{tabular}{c|ccccc} \hline
&&&&& \\
~~~~$C_n$~~~~ & ~~~~$\equ$~~~~ & ~~~~$\bar x_q$~~~~ & ~~~~$\egl$~~~~ &
~~~~$\bar x_g$~~~~ & ~~~~$<n_g>$~~~~ \\
&&&&& \\ \hline
&&&&& \\
$1/n$ & 0 & 0 & 1 & 0 & $\infty$ \\
&&&&& \\
$1/n^2$ & 1/2 & 1/6 & 1/2 & 0 & $\infty$ \\
&&&&& \\
$1/n^3$ & 2/3 & 2/9 & 1/3 & 1/9 & 3 \\
&&&&& \\
$1/n^4$ & 3/4 & 1/4 & 1/4 & 1/6 & 3/2 \\
&&&&& \\ \hline
\end{tabular}
\end{center}
\end{document}